\documentclass{aastex63}
\usepackage{hyperref}
\hypersetup{breaklinks=true}
\usepackage{breakurl}
\usepackage{amsmath,amssymb}
\begin{document}
\sloppy
\title{`Oumuamua is not Artificial}
\author{J. I. Katz}
\affiliation{Department of Physics and McDonnell Center for the Space Sciences\\
Washington University, St. Louis, Mo. 63130}
\email{katz@wuphys.wustl.edu}
\begin{abstract}
	I summarize evidence against the hypothesis that `Oumuamua is the
	artificial creation of an advanced civilization.  An appendix
	discusses the flaws and inconsistencies of the ``Breakthrough''
	proposal for laser acceleration of spacecraft to semi-relativistic
	speeds.  Reality is much more challenging, and interesting.
\end{abstract}
\section{Introduction}
Recently, the popular press \citep{NewYorker,NYT} has discussed the
hypothesis, promoted in a popular book \citep{L21}, that `Oumuamua, the
interstellar object that transited the Solar System in 2017 \citep{oumuamua},
is the product of an ``alien'' civilization, presumably reconnoitering the
Solar System, rather than a natural fragment (a ``Jurad'', an asteroid or
comet nucleus) that escaped from an extra-Solar planetary system
\citep{HZ17}.  There are several reasons why the alien civilization
hypothesis is not credible \citep{ISSI,Z21}:
\section{Rate}
The Panstarrs system detected one such object in its few years of operation.
A second interstellar object, clearly cometary, has also been detected.
There is some controversy over whether `Oumuamua is asteroidal or cometary:
imaging imposed very strict upper limits on its rate of outgassing.
\citet{M18} claimed non-gravitational acceleration but this was disputed by
\citet{K19}.  \citet{L21} attributed such an acceleration to the effect of
Solar radiation or the Solar wind on an artificial structure with a low
ballistic coefficient.  However, \citet{BL18} found that the reported
acceleration would imply a ballistic coefficient $\sim 0.1$ g/cm$^2$,
orders of magnitude greater than those considered for Solar sails or for
laser-accelerated spacecraft \citep{breakthrough}.

`Oumuamua had a velocity, far from the Solar System but with respect to it,
of about 26 km/s.  The smallest credible distance of a sending civilization
is about 10 light years (this volume contains 10--20 stars, enough that if
all the possible optimistic assumptions are made it might contain an
advanced civilization; the closest extra-Solar star is about 4 light years
away).  The transit time from that distance would be about $10^5$
years\footnote{The orbit of `Oumuamua has been extrapolated backward and
does not approach any star and possible planetary system for much
longer \citep{GWK17}, but we must allow for the possibility of a
non-ballistic trajectory powered by an on-board ion engine and nuclear
reactor.}.

A decision to launch toward our Solar System must have been made $\sim 10^5$
years ago.  Even if one imagines the launching civilization capable of
arbitrarily good observations, and able to infer from the state of the Earth
then that a technological civilization worth monitoring, alerting or
contacting would evolve in $\sim 10^5$ years, it could not predict the
chronology of Panstarrs or a similar system accurately.  Hence, unless it
(or we) have been unusually lucky, they must have launched $\gg 10^4$ such
probes.  This would be extraordinarily inefficient.
\section{Why Flyby?}
A flyby is an inefficient way to collect data and an unreliable way to 
attract notice.  As our space programs know, if you want to collect
information about a body, send an orbiter or a lander.  There is no
compelling argument against the presence of artificial orbiters in the Solar
System, or against landers on any body other than the Earth.  `Oumuamua was
neither.
\section{Tumbling}
`Oumuamua tumbled: its light curve was not periodic.  This is unusual, but
not unprecedented, among asteroids.  Internal friction damps tumbling, but
slowly in rocky material at low asteroidal temperatures and even more slowly
at the interstellar temperatures to which `Oumuamua was long exposed.  A
thin sail must be composed of flexible material or hinged (otherwise, it
could not be launched and deployed later).  Such materials have large
internal friction, as do hinges, and any tumbling might be damped rapidly
(although we cannot be confident of the properties of a material to be
invented by an advanced civilization).
\section{Further Observations}
\citet{L21} suggests a network of space telescopes to image any future
interstellar transiters (and more ordinary Solar System objects!).  A
diffraction-limited 10 m aperture has a resolution of about 50 nrad in blue
light.  At a distance of 1 AU that corresponds to a resolution of about
8 km, far too poor to resolve an object hundreds of meters across like
`Oumuamua.  Interferometry can do better, of course, but we cannot be
confident of an unambiguous distinction between the visibility functions of
natural and hypothetical artificial interstellar objects.

It is technically feasible to send probes to fly by such intruders.  
Panstarrs detects roughly one per year (and a future system likely many
more).  Velocity increments of 10 km/s, achievable in minutes with chemical
propellants (slow acceleration by ion engines would not be sufficient) would
enable interceptors stored in Solar orbits to fly by a significant fraction
of interstellar intruders.  Close-up imaging would be possible, or even
collision whose debris could be analyzed spectroscopically from Earth.
\section{Conclusion}
The hypothesis that `Oumuamua is the product of an advanced civilization
does not resolve any previously inexplicable conundrum, the necessary
justification for a speculative hypothesis.  `Oumuamua is entirely
explicable as a fragment expelled from its parent planetary system by
gravitational interaction \citep{HZ17}, at any time in the history of the
Galaxy.

The ``Breakthrough'' project has argued that it is feasible to accelerate,
using lasers, spacecraft of low ballistic coefficient (sails) to
semi-relativistic speed, orders of magnitude greater than the observed speed
of `Oumuamua; this is discussed in the Appendix.
\newpage
\appendix
\section{Laser-Accelerated Spacecraft}
The Breakthough project \citep{breakthrough} has suggested that it may be
possible, using the radiation pressure of a laser, to accelerate a low mass
spacecraft to semi-relativistic speeds for the purpose of interstellar
reconnaissance, by our future or a hypothetical ``alien'' civilization.
Representative parameters are a compact payload mass $M_p = 1$ g with 300
\AA\ thick Al foil sails of area $A = 10$ m$^2$, accelerated to $v \approx
0.2 c$ in a distance $d = 2.7 \times 10^{11}$ cm.  The implied uniform
acceleration $a = v^2/2d \approx 7 \times 10^7$ cm/s$^2$.  The sails have an
areal mass loading (ballistic coefficient) of $\approx 10^{-5}$ g/cm$^2$ and
a mass of $m = 1$ g; for the same $M_p$, larger sails produce only marginal
increases in performance.

The system has been modeled in more detail by \citet{P18}, and stability
issues addressed by \citet{ML17}.  Here I discuss a number of problems that
appear insuperable.
\section{Insuperable Problems}
\subsection{Energetics}
A $M_p = 1$ g spacecraft traveling at $0.2 c$ has a kinetic energy of $1.8
\times 10^{19}$ ergs.  The required laser power $P = M_p ac/2 \approx 100$
GW.  The required acceleration time $t = v/a = M_p cv/2P \approx 90$s.   The
distance traveled during acceleration $d = (M_p c/4P) v^2 \approx 2.7 \times
10^{11}$ cm.
\subsection{Focusing}
The proposed system must focus light at $2.7 \times 10^{11}$ cm onto a 3 m
sail.  For $1 \mu$ laser light, even diffraction-limited optics must have an
aperture of diameter $\gtrsim 1000$ m.  This may not be beyond the
capability of a hypothetical ``advanced civilization''.  More rapid
acceleration (\citet{breakthrough} suggest $d = 10^{11}$ cm, inconsistent
with their other parameters) would reduce the demands on the focusing optics
but would increase the required laser power and exacerbate the thermal
problems discussed in Sec.~\ref{vap}.
\subsection{Laser power}
Power engineers (quite apart from laser designers) will find the
requirement of delivering $100/\epsilon$ GW, where $\epsilon$ is the laser
efficiency, formidable.  It would require use of a large portion of the
present US electrical generating capacity for the duration of the
acceleration to accelerate even one 1 g spacecraft at a time, and sequential
acceleration of multiple spacecraft would multiply the required duration.
Some of these issues were discussed by \citet{L20}.
\subsection{The Sail Will Vaporize}
\label{vap}
\subsubsection{Absorbed energy}
The energy incident on the sail in order to accelerate a 1 g spacecraft is
$Pt$ or $9 \times 10^{19}$ ergs.  This should be compared to the latent heat
of melting of 1 g of aluminum of $\sim 10^{10}$ ergs.  If even $10^{-10}$ of
the incident laser energy is absorbed and not radiated, the sail melts.
This estimate is pessimistic because it ignores radiative cooling of the
sail, though metals are poor radiators for precisely the reason they are
good reflectors, but it indicates a formidable problem.  No metal has a
reflectivity higher than 0.9999 even for microwaves, and more typical values
for visible and near-visible light are $< 0.99$.  In other words, 100 GW
over 10 m$^2$ is 1 MW/cm$^2$, about $10^7$ times the intensity of sunlight,
and several orders of magnitude greater than intensities known to vaporize
metallic surfaces.
\subsubsection{Radiative cooling}
General materials issues are discussed by \citet{A18}.  Including radiative
cooling, in steady state (rapidly achieved for a very thin sail) the sail's
temperature would be
\begin{equation}
	T = T_{bb} (\epsilon_{laser}/\epsilon_{th})^{1/4}
\end{equation}
where the gray-body temperature $T_{bb} = [P/(A\sigma_{SB})]^{1/4} = 20000$
K, $A$ is the sail area (10 m$^2$), $\sigma_{SB}$ is the Stefan-Boltzmann
constant, $\epsilon_{laser}$ is the sail's absorptivity (equal to its
emissivity) at the laser wavelength ($1\mu$) and $\epsilon_{th}$ is the
sail's emissivity at MWIR and LWIR wavelengths corresponding to its
equilibrium temperature.  The ratio of emissivities is greater than unity
because metals are better reflectors (and worse emitters) at the longer
wavelengths of their thermal emission and because very thin layers (of any
material) have smaller emissivities than thick layers.  Thermal emission by
a metallic sail cannot prevent it from melting unless $\epsilon_{laser}/
\epsilon_{th} \lesssim 10^{-5}$; even if emitting as a black body, this would
require $\epsilon_{laser} \lesssim 10^{-5}$, about three orders of magnitude
less than the actual intrinsic emissivities and absorptivities of metals.

Dielectrics may have very low absorptivity \citep{A18} because their
electrical conductivity, the classical source of absorption in metals, may
be essentially zero.  However, the required extremely low absorptivity has
not been demonstrated.  Real materials do not have their ideal properties;
even with zero electrical conductivity, they are likely to contain absorbing
impurities or ``dirt''.  Dielectric films are less reflective than metals
because the magnitudes of their real dielectric constants are much less than
the magnitudes of the imaginary dielectric constants of metals.  Unlike
metals, dielectrics cannot be rolled into the ultra-thin ($\sim 300$ \AA )
sheets required by Breakthrough Starshot.
\subsubsection{Better radiator?}
Thermal problems might be mitigated if the metal were backed with a thin
layer of an effective (necessarily insulating) radiator.  Supposing a
black radiator (overoptimistic, because thin sheets have lower emissivity
than half-spaces) and (also overoptimistic) $\epsilon_{laser} = 10^{-4}$
yields $T = 2050$ K, still exceeding the melting point of Al of 933 K.  Each
of these assumptions is unrealistic (the actual $\epsilon$ at $\lambda =
1\mu$ of Au is 0.03 and that of Al 0.07), and ignores the difficulty of
fabrication.  Another optimistic calculation \citep{K17} implies a minimum
reflectivity $> 0.999994$.  This also exceeds any known value.
\subsection{Utility?} 
How does a 1 g spacecraft carry any useful sensors?  How can it determine
what is around $\alpha$ Centauri (presumably the interest is in habitable
planets)?
\subsection{Data return?}
The issues of returning data from interstellar distances have been discussed
by \citet{H19,P20}.  The difficulties are parfticularly severe for the
proposed 1 g spacecraft at a distance of $4 \times 10^{18}$ cm.  Such a low
mass spacecraft would have very little power available to radiate, and the
gain, that might compensate for limited power, of a small antenna would be
small.
\subsubsection{Signal strength}
A half-wave dipole antenna, using a hypothetical 100 W of power obtained
from photocells illuminated by the parent star of the planet studied, would
produce $10^{-36}$ W/cm$^2$ at the Earth at 1 pc distance, or $10^{-26}$ W
in a 1 km$^{2}$ array.  For a bandwidth $B$ the corresponding temperature
would be $\sim 1 (\text{1 Hz}/B)$ mK, that may be compared to
present-state-of-the-art antenna temperatures $\approx 20$ K.  Extensive
coherent integration would be required to discriminate the signal from the
incoherent thermal cosmic 3 K background and the thermal emission of the
parent star.  A high gain transmitting antenna would improve these values,
but would be difficult to include within a 1 g mass budget.
\subsubsection{Visible light?}
An optical system could collimate the beam, but at the price of quantum
noise more than $10^6$ times greater than at 100 MHz, and folding such
optics into a 1 g spacecraft would be a formidable problem.  A
diffraction-limited 1 m aperture would increase the received intensity by
$10^{12}$ compared to a dipole, or to $10^{-18}$ W in a 10 m receiving
telescope, or 3 photons/s.  It is difficult enough to detect planets,
reflecting $10^{17}$ W of visible light (giving the same brightness as the
hypothetical diffraction-limited aperture transmitting $10^5$ W) near stars.
\subsection{Relays}
The total power required to transmit a signal may be reduced by a factor of
about $N$ if a relay chain of $N$ equally spaced repeaters is established
along the transmission path; the power at each repeater is reduced by a
factor $\sim N^2$ (by the inverse square law).  These repeaters would be in
deep space where the only possible power source would be nuclear reactors,
that must have minimimum masses of tens of kg each in order to achieve
criticality.  Decay of isotopes like $^{90}$Sr and $^{137}$Cs, with
half-lives of decades, produces $\sim 1$ W/g, that can only be converted to
electricity with an efficiency of a few percent.  To be useful, this massive
relay chain would have to be deployed behind the high-velocity probe at
speeds approaching that of the probe; deployment at conventional spacecraft
speeds $\sim 30$ km/s would require tens of thousands of years to reach the
nearest stars at distances of $\sim 1$ pc.
\section{Not Self-Consistent}
The force required to accelerate the $M_p = 1$ g payload $F = M_p a = 2P/c
= 7 \times 10^7$ dynes.  This must be transmitted from the sail to the
compact payload.  The payload dimensions are not specified; consider an
integrated circuit with area 10 cm$^2$, that may be described as a circular
disc of radius 1.7 cm and circumference $C = 10$ cm.  The stress in the sail
where it is attached to the payload is $F/Ch$, where $h$ is the sail's
thickness.  For the nominal $h = h_0 = 300$ \AA, the thickness of an Al sail
that can produce the assumed acceleration with the assumed laser intensity,
the stress is $> 2000$ kbar, $> 600$ times the strength $\sigma \approx 3
\times 10^9$ dynes/cm$^2$ of aluminum.  The sail must be thicker.

The sail has its minimum mass if its thickness $h(r) = F/(2 \pi r \sigma)$.
Then the mass of the thickened ($h > h_0$) region of the sail
\begin{equation}
	M_s = \int\! 2 \pi r h(r) \rho\,dr = FR{\rho \over \sigma} = M_p
	{R \over R_c},
\end{equation}
where $R = F/(2 \pi \sigma h_0)$ is the radius at which $h$ decreases to its
minimum value $h_0$ and the characteristic radius $R_c \equiv a \sigma/\rho
\approx 14$ cm.  Sail material at $r > R$ does not contribute significantly
to accelerating the payload and its stress is $< \sigma$.  Then
\begin{equation}
	M_s = {(M_p a)^2 \rho \over 2 \pi h_0 \sigma^2}.
\end{equation}
For the assumed $M_p$ and $a$, $R \approx 14$ m, $A \approx 600$ m$^2$ and
$M_s \approx 100$ g.  The proposed $A = 10$ m$^2$ ($R = 1.7$ m) sail must
everywhere be thicker than the assumed 300 \AA\ in order to carry the
necessary force, and would have a mass of 14 g.  Neither of these more
massive sails would provide the assumed acceleration with the assumed laser
power.

To reduce the sail mass to that of the payload would require $R \le R_c$.
The sail's area would be $\le 0.06$ m$^2$, much less than the assumed 10
m$^2$.  The laser intensity would be at least 170 times greater than that
previously assumed, that would already exceed the tolerable heat load on the
sail by orders of magnitude.  The proposed design is not self-consistent for
any payload mass $M_p > 0.01$ g if the sail is made of Al.  It might be
self-consistent if the sail were made of aramid or ultra high molecular
weight polyethylene (UHMWPE) with $\sigma = 3 \times 10^{10}$ dyne/cm$^2$,
if these materials could be made in sufficiently thin (300 \AA) sheets, were
sufficiently reflective, and absorbed only an infinitesimal fraction of the
incident laser light.

\end{document}